\documentclass{article}
\usepackage{color}
\usepackage{spconf,amsmath,graphicx}
\usepackage{comment}
\usepackage{blindtext}

\title{DEEP CONVOLUTIONAL AND RECURRENT NETWORKS FOR POLYPHONIC INSTRUMENT CLASSIFICATION FROM MONOPHONIC RAW AUDIO WAVEFORMS}

\name{Kleanthis Avramidis$^*$\textsuperscript{1,2},
Agelos Kratimenos$^*$\textsuperscript{1,2},
Christos Garoufis\textsuperscript{1,2}, \vspace{-0.4cm}
\thanks{$^*$The first two authors contributed equally.}}

\address{\textsuperscript{1} School of ECE, National Technical University of Athens, 15773 Athens, Greece \\
\textsuperscript{2} Robot Perception and Interaction Unit, Athena Research Center, 15125 Maroussi, Greece \\
\small{kle.avramidis@gmail.com, ageloskrat@yahoo.gr, cgaroufis@mail.ntua.gr, \{nzlat, maragos\}@cs.ntua.gr}}

\begin{document}

\ninept
\maketitle

\begin{abstract}
Sound Event Detection and Audio Classification tasks are traditionally addressed through time-frequency representations of audio signals such as spectrograms. However, the emergence of deep neural networks as efficient feature extractors has enabled the direct use of audio signals for classification purposes.
In this paper, we attempt to recognize musical instruments in polyphonic audio by only feeding their raw waveforms into deep learning models. Various recurrent and convolutional architectures incorporating residual connections are examined and parameterized in order to build end-to-end classifiers with low computational cost and only minimal preprocessing. We obtain competitive classification scores and useful instrument-wise insight through the IRMAS test set, utilizing a parallel CNN-BiGRU model with multiple residual connections, while maintaining a significantly reduced number of trainable parameters.
\end{abstract}
\begin{keywords}
Raw Waveforms, End-to-End Learning, Polyphonic Music, Instrument Classification
\end{keywords}

\section{Introduction}

Waveforms are abstract representations of sound waves and, when recorded, they constitute convoluted signals that incorporate noise from the complexity of the recorded sound event, the acoustic scene, as well as the recording equipment. Complex sound events such as spoken dialogues or simultaneously playing musical instruments (i.e. polyphonic music) can be challenging in extracting meaningful information. Thus, audio classification tasks traditionally discard direct waveform modeling in favor of richer time-frequency feature representations \cite{time-freq}. In fields like Speech Recognition \cite{speech} and Music Information Retrieval \cite{DBLP:conf/ismir/GururaniSL19}, such methods take advantage of the discriminative information of the signals' spectra, which is aligned to the human auditory system.

In Instrument Classification particularly there is strong intuition into utilizing frequency-related representations, since musical notes and instruments are densely associated with specific frequency events. Thus, most research works in the field incorporate spectrograms in their analysis. It is however challenging and computationally expensive to design specialized feature representations for each different recognition task, especially when contemporary deep learning models emerge as strong feature extractors for end-to-end classification. In this paper we address this challenge by parameterizing deep recurrent and convolutional networks to model raw audio waveforms efficiently. Our analysis is concentrated on handling the high input dimensionality of the waveforms, mining temporal features and preserving their low-level spatial locality, while reducing the computational cost of the process. We propose a lightweight end-to-end classifier for Instrument Classification that shows comparable performance to state-of-the-art spectrogram-based architectures, including our previous work on the task \cite{kratimenos2020}.

The rest of the paper is organized as follows: Sec.~2 provides a review of related research in Audio Signal Processing using raw waveforms, as well as Instrument Classification. The architectures that are used throughout our experiments are analyzed in Sec.~3. Sec.~4 describes the experimental setup, the dataset and the evaluation methods to be followed, whereas in Sec.~5 we discuss the results of our experiments. Finally, in Sec.~6 we present our conclusions as well as propose further directions for future work.

\begin{figure}[t!]
    \centering
    \includegraphics[scale=0.51]{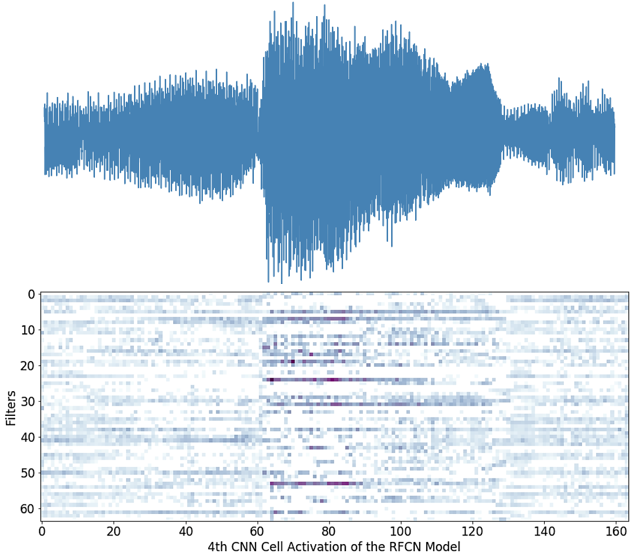}
    \caption{Intermediate activation of the Residual FCN Model (Sec.~3.2) for the above 
    1-sec piano sample.
    }
    \vspace{-0.3cm}
\end{figure}

\section{Related Work}
\label{sec:format}

\begin{figure*}[t!]
    \centering
    \includegraphics[scale=0.6]{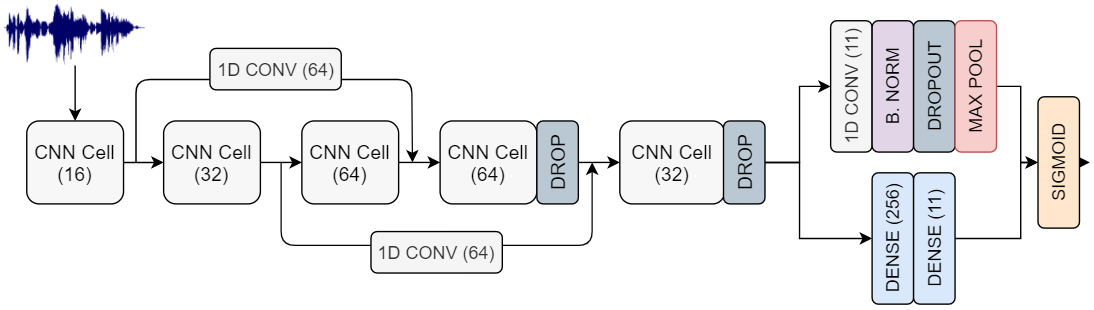}
    \caption{The DCNN, FCN and RFCN architectures used in the experimental evaluation.
    }
    \vspace{-0.2cm}
\end{figure*}

Deep neural networks, which have achieved state-of-the-art performance in audio recognition \cite{bae2016acoustic} by operating directly in the time domain, have blurred the line between representation learning and predictive modeling. Convolutional networks particularly have shown competitive performance, in some cases matching that of classical spectrogram-based models \cite{Sainath2015}. In speech analysis and synthesis, WaveNet \cite{vandenoord2016} is a benchmark model fed with audio waveforms, whereas \cite{chorowski2019} achieves robust representation learning by utilizing WaveNet Autoencoders that use waveforms directly as input. In Music Information Retrieval, a number of works have attempted to acquire high-level features like melody and pitch \cite{crepe}, while waveform-based architectures have also recorded competitive results in music \cite{stoller2018} and speech \cite{luo2019} source separation.

As far as Instrument Classification is concerned, until recently the majority of works utilized time-frequency representations and datasets of solo recordings or excerpt-level annotations (e.g. IRMAS \cite{Bosch2012ACO}, MedleyDB \cite{Bittner2014MedleyDBAM}). While traditional research, partially due to the challenge of labeling multi-instrumental music, focused upon monophonic audio \cite{herrera2003automatic}, recent studies address polyphonic tasks, relying on the efficiency of deep learning models. Specific points of focus include investigation of the optimal input temporal resolution \cite{Gururani2018InstrumentAD, Han2017DeepCN} and the design of the convolutional filters involved \cite{Pons2017TimbreAO}, while we have also experimented with sophisticated augmentation methods, attempting to isolate timbre-like characteristics \cite{kratimenos2020}.

\section{Architectures}

\subsection{Recurrent Networks}

Recurrent neural networks (RNN) have been widely used in waveform modeling and classification, thanks to their ability to model long-range temporal dependencies. In our baseline network we employ the Bidirectional Gated Recurrent Unit (BiGRU). GRU architectures have shown comparable performance to Long Short-Term Memory (LSTM) units in processing audio sequences \cite{gru}, although they inherit a less complex structure and provide lower computational costs. Moreover, Bidirectional units consider both past and future audio context, which intuitively assists our task.

\begin{table}[hbt!]
\centering
\begin{tabular}{|c|c|}
\hline
\textbf{Number of Layers} & \textbf{Number of Units} \\ \hline
1    & 128 or 256           \\
2    & 128, 64   \\ \hline
\multicolumn{2}{|c|}{Dropout (0.5)} \\ \hline
\multicolumn{2}{|c|}{Output Dense}         \\ \hline
\end{tabular}
\caption{BiGRU Architecture Configurations.}
\end{table}

\begin{figure}[t!]
    \centering
    \includegraphics[scale=0.4]{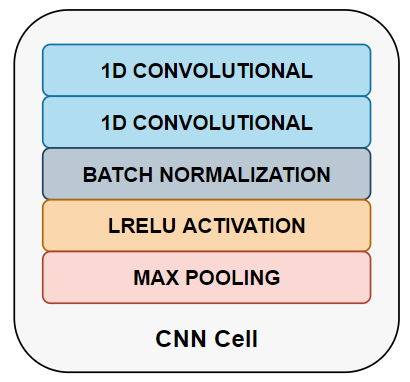}
    \caption{The utilized CNN cell structure.}
    \vspace{-0.3cm}
\end{figure}

Regarding this recurrent module, we experiment with the optimal number of layers and utilized units. Specifically, we trained 3 different networks: one with 1 BiGRU layer of 128 units, one with 1 BiGRU layer of 256 units and one with 2 BiGRU layers of 128 and 64 units respectively, as shown in Table~1. We used a fully connected layer to produce the models' output as well as a Dropout layer right before, to enhance the models' ability to generalize. In order to effectively reduce the high dimensionality of the waveforms we apply max pooling at the input level, with a pool size of 3 samples. 

\subsection{Convolutional Networks}

Convolutional Neural Networks (CNN) traditionally operate on images \cite{Simonyan2015} or, in Audio Classification, on time-frequency representations like spectrograms \cite{Gururani2018InstrumentAD}, being an extremely coherent feature extraction method for Deep Learning. 
However, they can provide useful results when applied to 1D signals as well \cite{Dai2017}. We base our CNN on the architecture used in \cite{kratimenos2020} that yielded strong results on the IRMAS Dataset. In order to adjust to the different input form, we substitute 2D with 1D Convolutional layers and fine-tune the pooling parameters to handle the high input dimensionality. The CNN cell structure is composed of 2 stacked convolutional layers with the same number of filters, a Batch Normalization layer that enhances the training efficiency \cite{pmlr-v37-ioffe15}, a Leaky ReLU activation and a max pooling layer with kernels of varied length (Fig.~3).

We place 5 CNN cells in a row with 16, 32, 64, 64 and 32 filters respectively, using kernel and pooling sizes equal to 3 for the first three layers and 5 for the rest. This module is followed by two fully connected layers (denoted as DCNN), something that increases substantially the number of its trainable parameters. We thus experiment by removing all the fully connected layers to form a Fully Convolutional Network (FCN). The 11D output vector is then estimated through an additional convolutional layer of unit kernel, followed by Global Average Pooling. The adoption of FCNs in modeling raw audio waveforms sharply reduces the number of parameters that we need to train, while it can force the network to learn meaningful features in its hidden layers, that keep their temporal locality throughout the architecture \cite{Dai2017}. The final configuration is a Residual FCN (RFCN), where we simply embed skip connections to the previous model, as shown in Fig.~2. Through residual connections the model is able to propagate low-level features throughout the network.

\subsection{Combined Networks}

The utilized feature extraction and prediction methods are capable of learning different types of features. It has been demonstrated \cite{bae2016acoustic} that convolutional nets concentrate on spatial features and, in the context of waveforms, on temporally local correlations, while recurrent ones are useful in modeling longer-term temporal structure. We can therefore expect that a combined Convolutional-Recurrent Neural Network (CRNN) would further enhance the performance of the highlighted architectures, so we attach the best performing BiGRU model into our RFCN model. In order to keep the temporal resolution of the RNN to a feasible magnitude, we experiment by embedding it after the 2nd, the 3rd and the 4th CNN cell, as well as by feeding the CNN output to the recurrent units. Specifically, the embedded model takes the output of the corresponding CNN cell and its output is reduced to the number of classes through additional convolution with a unit kernel, followed by Global Average Pooling. The two 11D vectors are then averaged before the Sigmoid activation. In this way we empirically search the optimal way of integrating the recurrent model information into a robust classifier.

\begin{table}[t!]
\scalebox{0.9}{
\begin{tabular}{c|c|c|c||c}
\textbf{BiGRU} & \textbf{F1-micro \%} & \textbf{F1-macro \%} & \textbf{LRAP \%}  & \textbf{\#Params} \\ \hline

1 (128)    
           &   43.76 $\pm$ 1.95
           &   37.37 $\pm$ 1.90
           &   57.26 $\pm$ 3.28
           &   \textbf{103.4K} \\
1 (256)  
           & 43.51 $\pm$ 2.46   
           &   39.19 $\pm$ 2.23
           &   58.47 $\pm$ 2.73
           &   403.4K\\
           
2     & \textbf{49.28 $\pm$ 2.45}
      & \textbf{43.18 $\pm$ 3.11}
      & \textbf{67.07 $\pm$ 1.81}
      & 225.6K\\
           
\end{tabular}
}
\caption{Results for the Recurrent Networks discussed in Sec.~3.1, subject to the number of GRU layers and the  number of units.}
\end{table}

\begin{table}[!t]
\scalebox{0.9}{
\begin{tabular}{c|c|c|c||c}
\textbf{Models}  & \textbf{F1-micro \%} & \textbf{F1-macro \%} & \textbf{LRAP \%} & \textbf{\#Params} \\ \hline

DCNN    
        &  55.32 $\pm$ 0.55
        &  48.30 $\pm$ 0.31
        &  73.48 $\pm$ 0.38
        &  1.14M \\
FCN     
        &  58.45 $\pm$ 0.36
        &  49.96 $\pm$ 0.29
        &  75.13 $\pm$ 0.32
        &  \textbf{81.8K} \\
RFCN    
        &  \textbf{58.55 $\pm$ 0.22}
        &  \textbf{50.22 $\pm$ 0.35}
        &  \textbf{75.14 $\pm$ 0.23}
        & 85K \\

\end{tabular}
}
\caption{Results for the Dense Connected Neural Network (DCNN), the Fully Connected Network (FCN) and the Residual FCN (RFCN) discussed in Sec.~3.2.}
\vspace{-0.2cm}
\end{table}

\section{Experimental Setup}

\subsection{Dataset \& Training}

The IRMAS dataset \cite{Bosch2012ACO} is used to train and test our models, as it has been extensively researched for the task of Instrument Classification. IRMAS is divided into a training set containing 6705 audio segments of 3 seconds each and a testing set containing polyphonic tracks of various lengths. Each of the 3-sec training snippets is annotated with exactly 1 out of the 11 available predominant instruments, while the polyphonic tracks in the testing set contain 2--4 instruments. 
We choose to cut each track into 1-sec segments, since this temporal resolution increases the data volume and helps the model generalize better, as indicated in \cite{Gururani2018InstrumentAD, Han2017DeepCN}. Each waveform is then downsampled to 22.05 kHz, downmixed to mono and normalized by its root-mean-square energy. Since we are interested in classifying only raw waveforms, no further pre-processing is applied.

The training data are then partitioned into 5 subsets to perform cross-validation for each of the above-mentioned architectures. All networks were trained using binary cross-entropy loss since the task is modeled as multi-class (11 instruments) and multi-label (instruments can co-play). Adam optimizer \cite{kingma14} is used to optimize the loss function, with an initial learning rate of 0.001 and 10\% decay rate per 4 epochs of non-decreasing validation loss. The batch size is set to 64 after fine-tuning. We also perform Early Stopping by monitoring the validation loss with a patience of 7 epochs.

\subsection{Evaluation Metrics}

Each model is evaluated at the IRMAS test set, consisting of 2355 polyphonic music tracks, ranging from 5 to 20 sec duration. During the evaluation process, we partition each track into 1-sec segments, compute the frame-level predictions and then average them in order to extract a single track prediction. This method will produce reliable results because each labeled instrument is always active for the whole duration of the track. For the particular polyphonic classification we utilize two metrics. The first one is the F1 Score,  which is widely used in many relative studies \cite{Bosch2012ACO, Han2017DeepCN} and provides a balanced view of multi-class performance. In order to calculate an overall score, we compute the average of the per-instrument scores at both micro and macro scales. The second one is Label Ranking Average Precision (LRAP), a rank-based metric proposed in \cite{ss-bbstc-00}. LRAP is suitable for multi-label evaluation as it is threshold-independent and measures the classifier’s ability to assign higher scores to the correct labels associated to each sample.

\begin{table}[!t]
\scalebox{0.9}{
\begin{tabular}{c|c|c|c||c}
\textbf{Models} & \textbf{F1-micro \%} & \textbf{F1-macro \%} & \textbf{LRAP \%} & \textbf{\#Params} \\ \hline


CRNN$_2$ 
         & 59.80 $\pm$ 0.66
         & 53.20 $\pm$ 0.52
         & 74.16 $\pm$ 0.66
         &  \textbf{1.03M}\\
CRNN$_3$ 
         &  \textbf{60.77 $\pm$ 0.26}
         &  \textbf{54.31 $\pm$ 0.35}
         &  \textbf{74.74 $\pm$ 0.39}
         &  1.07M\\
CRNN$_4$ 
         & 60.07 $\pm$ 0.67
         & 53.73 $\pm$ 0.59
         & 74.11 $\pm$ 0.50
         & 1.08M \\
CRNN$_5$ 
         & 59.21 $\pm$ 0.56
         & 52.18 $\pm$ 0.46
         & 74.32 $\pm$ 0.65
         & 1.03M \\
         
\end{tabular} }
\caption{Results for the combined CNN and RNN networks discussed in Sec.~3.3. The subscript denotes the layer in which the latter was connected to the former.}
\vspace{-0.1cm}
\end{table}

\section{Results and Discussion}

\subsection{Architecture Comparison}

Table 2 shows the accuracy scores for the recurrent neural models proposed in Sec.~3.1. It is clear that a simple recurrent network cannot sufficiently decode the information included in a waveform. Still, the best model emerges from a combination of two BiGRU layers with 128 and 64 units, respectively, which by far outperforms the single GRU layer. Further experiments show that adding a CNN Cell before the recurrent units significantly improves the classification, indicating the efficiency of the models described in Sec.~3.3. \par

\begin{figure*}[t!]
    \centering
    \includegraphics[scale=0.6]{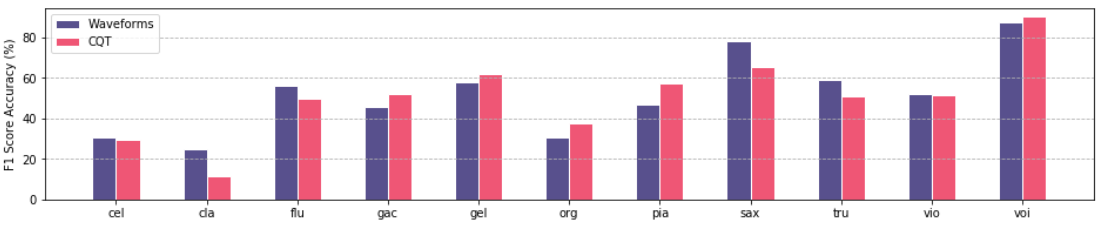}
    \caption{Instrument-wise performance of the proposed model and the monophonic \cite{kratimenos2020} in terms of F1-score.}
\end{figure*}

On the other hand, as we see in Table~3, 1D convolutional models are capable of extracting the most discriminative features from raw waveforms, almost as well as 2D convolutional models that work on spectrogram inputs \cite{Han2017DeepCN, kratimenos2020}. Furthermore, removing the dense layers not only reduces the number of model trainable parameters, and thus the training time, but also increases the accuracy substantially. We argue that, in the absence of a dense layer, the network generalizes better upon the information from the spatial processing. Additionally, connecting the output of the first CNN cell with the third one, and the output of the second CNN cell with the fourth, slightly increases the results with no additional cost in parameters or training time. Other types of residual connections though do not seem to yield consistently improved performance.

\begin{table}[t!]
\scalebox{0.89}{
\begin{tabular}{c|c|c|c||c}
\textbf{Models} & \textbf{F1-micro} & \textbf{F1-macro} & \textbf{LRAP} & \textbf{\#Params} \\ \hline
Bosch et al. \cite{Bosch2012ACO}   &   0.503  &  0.432  &  --   & --\\
Pons et al. \cite{Pons2017TimbreAO}      &   0.589  &  0.516  & --   & -- \\
Han et al. \cite{Han2017DeepCN}       &   0.602  &  0.503  & --   & --\\
Kratimenos et al. \cite{kratimenos2020} &  \textbf{0.616} & 0.506 & \textbf{0.767} & 24.3M \\
Proposed                    &   0.608  & \textbf{0.543} & 0.747 & \textbf{1.07M} \\

\end{tabular} }
\caption{Comparison of our work with previous performances on the IRMAS Dataset}
\vspace{-0.5cm}
\end{table}

To optimally combine the temporal information extracted from the RNN and the already utilized spatial characteristics drawn from the residual network, we additionally assess the performance of the combined networks described in Sec.~3.3. Simply averaging the RNN and CNN model outputs lowers though the classification accuracy, something we attribute to the inadequate standalone performance of the BiGRU (see Table 2). We thus inserted the RNN model in various locations in the RFCN architecture that yielded optimal accuracy. From the scores we report at Table~4, we notice that there is no observed improvement in the model performance as far as the LRAP metric is concerned. However, there is a steady increase at F1 scores, about 2\% and 4\% at micro and macro scales, respectively. It should be mentioned at this point that the combined models consist of a significantly larger number of parameters than the fully convolutional ones, while the DCNN is the model with the most parameters, despite its lower performance.

The optimal architecture is constructed by embedding the BiGRU module after the 3rd CNN cell of the residual fully convolutional model and by averaging the 2 models' output in the end. The proposed model yields an LRAP score close to 75\% and F1 scores comparable to the literature \cite{Han2017DeepCN, kratimenos2020, Pons2017TimbreAO} on the IRMAS Dataset. Specifically, F1 micro surpasses most studies on the task, while we observe dominant performance at the more competitive F1 macro score, in which case the model even surpasses the performance achieved in our previous work with the use of complex data augmentation \cite{kratimenos2020}. Those results are obtained with a significantly reduced number of trainable parameters, low training and testing time and minimal pre-processing. In order to emphasize this, we attempt to train the model utilized in~\cite{kratimenos2020} by incorporating a significantly reduced number of parameters, without altering the architecture. Specifically, we cut the number of filters and the final dense layer of the model. The results (Table~6) show that this network falls behind the proposed one by 6\% and an average 8\%, regarding LRAP and F1 scores respectively.

\subsection{Instrument-wise Analysis}

To get a more thorough insight into the waveform characteristics of different instruments and how much these can assist the task of Instrument Classification, we examine the class-wise performance in terms of the F1 metric. The results are visualized in Fig.~4, along with the corresponding results obtained from Constant Q Transform (CQT) spectrogram modeling from our previous work~\cite{kratimenos2020}. We clearly observe that brass instruments are recognized much better using raw waveforms, compared to CQT representations. In specific, clarinet, flute, saxophone and trumpet achieve 14\%, 7\%, 13\% and 9\% increase respectively in F1-score when waveforms are considered. On the other hand, predominant instruments, i.e. electric/acoustic guitar, piano/organ and the human voice, are distinguished better through processing their CQT representation, with the highest difference observed on piano with 10\% change in F1-score. Apart from that, the absolute performance of the instruments pretty much resembles the findings of the CQT-based study.

\begin{table}[t!]
\scalebox{0.99}{
\begin{tabular}{c|c|c|c||c}
\textbf{Models} & \textbf{F1-micro} & \textbf{F1-macro} & \textbf{LRAP} & \textbf{\#Params} \\ \hline

Proposed
           &   \textbf{0.608}
           &   \textbf{0.543}
           &   \textbf{0.747}
           &   \textbf{1.07M} \\
\cite{kratimenos2020} Reduced&  0.520  
           &   0.458
           &   0.689
           &   1.20M\\

\end{tabular}
}
\caption{Comparison between the proposed architecture and the state-of-the-art after parameter reduction.}
\vspace{-0.1cm}
\end{table}

\section{Conclusions}

In this paper we attempt to perform polyphonic instrument classification from monophonic music data with the usage of their raw audio waveforms. We experiment with various architectures that are favourable towards waveforms, like Fully Convolutional and Residual Nets and we also attempt to embed a recurrent modules to the optimal Architecture so as to fuse their discriminating information. A residual FCN-BiGRU model with a total of 1 million parameters outperforms the state-of-the-art model, utilizing CQT spectrograms and holding 24 million parameters, in the F1-macro metric by 4\%, while it is comparable in the F1-micro and LRAP metrics. A more thorough experiment on the performance of each instrument independently shows that brass instruments are being identified easier through waveforms, while predominant instruments, like piano or the electric guitar, benefit more from time-frequency representations.

Future work should therefore deal with alternate methods to exploit a recurrent neural network when fed with raw waveforms, as well as with methods to enhance the recognition performance of predominant instruments. Finally, additional experiments should address how noise, that is highly incorporated within waveform signals, affects the model capabilities.

\nocite{*}
\bibliographystyle{IEEEbib}
\bibliography{refs}

\end{document}